\documentclass[epj,final]{svjour}
\usepackage{latexsym}
\usepackage{amsmath}
\usepackage{amssymb}
\usepackage{graphics}
\usepackage{cuted}
\usepackage{hyperref}
\begin{document}

\title{Pseudo-transition between antiferromagnetic and charge orders in a minimal spin-pseudospin model of one-dimensional cuprates}
\author{Jozef Stre\v{c}ka\inst{1} \fnmsep \thanks{\email{jozef.strecka@upjs.sk}} 
        \and Katar\'ina Karl'ov\'a\inst{2} \fnmsep \thanks{\email{katarina.karlova@cyu.fr}}}                     
\institute{Department of Theoretical Physics and Astrophysics, Faculty of Science, \\ 
P. J. \v{S}af\'{a}rik University, Park Angelinum 9, 040 01 Ko\v{s}ice, Slovak Republic \and 
Laboratoire de Physique Th\'eorique et Mod\'elisation, CY Cergy Paris Universit\'e, \\
Avenue A. Chauvin, Pontoise, 95302 Cergy-Pontoise, France}
\PACS{{05.50.+q}{Lattice theory and statistics}
 \and {75.10.-b}{General theory and models of magnetic ordering}
 \and {75.10.Pq}{Spin chain models}
 \and {75.30.Kz}{Magnetic phase boundaries}
 \and {75.40.Cx}{Static properties}}
\authorrunning{\textit{J. Stre\v{c}ka and K. Karl'ov\'a}}
\titlerunning{Pseudo-transition between antiferromagnetic and charge orders}
\date{Received: date / Revised version: date}
\abstract{This study rigorously investigates the phenomenon of a pseudo-transition in a minimal spin-pseudospin model, serving as a simplified model of one-dimensional cuprate chains [CuO]$_\infty$, by making use of the transfer-matrix method. The studied spin-pseudospin model of one-dimensional cuprate chains [CuO]$_\infty$ reveals a peculiar pseudo-transition between a charged-ordered phase and an antiferromagnetic phase, which is accompanied with anomalous behavior of magnetic and thermodynamic quantities. While the entropy and short-range correlations undergo near the pseudo-transition steep continuous changes reminiscent of a discontinuous phase transition, the specific heat displays a sizable sharp peak akin to a continuous phase transition.} 

\maketitle

\section{Introduction}\label{sec1}

One-dimensional lattice-statistical spin models traditionally attract a great deal of attention due to their conceptual simplicity and exact solvability \cite{mat93}. Previous exact solutions for one-dimensional spin models with short-range interactions and non-singular potential have consistently ruled out the existence of finite-temperature phase transitions in accordance with the non-existence theorems rigorously proved by van Hove \cite{hov50}, Cuesta and S\'anchez \cite{cue04}. However, recent studies revealed that a variety of enigmatic one-dimensional lattice-statistical models may display pseudo-transitions at nonzero temperatures, which closely resemble genuine thermal phase transitions. Notable examples of one-dimensional lattice-statistical models displaying thermal pseudo-transitions include the coupled spin-electron double-tetrahedral chain \cite{gal15}, spin-1/2 Ising-Heisenberg ladder \cite{roj16} and tube \cite{str16}, spin-1/2 Ising-XYZ diamond chain \cite{car18,car19}, spin-1/2 and mixed spin-(1/2, 1) Ising-Heisenberg double-tetrahedral chains \cite{ono20,roj20}, spin-1/2 Ising diamond and tetrahedral chain \cite{str20,nov20}, mixed spin-(1/2,1) Ising hexagonal nanowire \cite{pim22}, spin-1/2 Ising ladder with trimer rungs \cite{hut21,yin24}, as well as, one-dimensional q-state Potts models \cite{pan21,pan23}. Thermal pseudo-transitions of all aforementioned one-dimensional lattice-statistical models differ from genuine phase transitions in several important respects: first-order derivatives of a free energy display in vicinity of a pseudo-transition a steep but continuous change instead of a true discontinuity and second-order derivatives of a free energy exhibit in vicinity of a pseudo-transition a sizable finite peak instead of a power-law divergence \cite{sou18,oro20}. In spite of this fact, sharp (albeit finite) peaks in correlation length, specific heat and magnetic susceptibility closely mimic peculiar power-law divergences near pseudo-transitions, characterized by a set of universal quasi-critical exponents \cite{roj19,kro21}. Moreover, anomalous behavior of quantities observed around pseudo-transitions is accompanied by significant finite-size effects \cite{roj21}, whereby the pseudo-transitions can be approached arbitrarily closely \cite{yin24}.

In this paper, we introduce and exactly solve a minimal spin-pseudospin model for [CuO] centers of the polymeric cuprate chain [CuO]$_\infty$, which will extend the realm of one-dimensional lattice-statistical models displaying pseudo-transition. The simplified spin-pseudospin model of the one-dimensional cuprate chain [CuO]$_\infty$ will take into consideration three valence states of Cu$^{1+}$, Cu$^{2+}$ and Cu$^{3+}$ centers corresponding to the many-electron valence states of [CuO]$^{-}$, [CuO]$^{0}$ and [CuO]$^{+}$ centers as three components of the pseudospin triplet \cite{mos11,mos13,mos19}. The spin-pseudospin model of one-dimensional cuprate chain [CuO]$_\infty$ will take into account a mutual electrostatic correlation between non-magnetic valence states of [CuO]$^{-}$ and [CuO]$^{+}$ centers (nominally Cu$^{1+}$ and Cu$^{3+}$ centers), as well as, an antiferromagnetic superexchange interaction between the spin-1/2 Cu$^{2+}$ centers mediated by O$^{2-}$ anions within the electroneutral [CuO]$^{0}$ centers. The approach based on a minimal spin-pseudospin model has already demonstrated its utility for two-dimensional cuprates composed of CuO$_2$ planes, which exhibit rather complex phase diagrams due to competition between insulating antiferromagnetic, charged ordered and superconducting phases \cite{mos19,pan16,pan17,pan19,pan20,pan22}. Although the present spin-pseudospin model was primarily developed for the purpose of investigating a potential pseudo-transition between antiferromagnetic and charge ordered phases, it may also shed light on essential features of one-dimensional cuprates Sr$_2$CuO$_3$ \cite{ami95,mai97,sch12,sch18}, Ca$_2$CuO$_3$ \cite{yam95} or SrCuO$_2$ \cite{nag97,bou18} incorporating cuprate chains [CuO]$_\infty$ as their basic structural motif. 

The paper is structured as follows. The spin-pseudospin model will be introduced in Section \ref{sec2} along with a concise overview of the method used for its exact treatment. The discussion of the most interesting results is presented in Section \ref{sec3} and finally, some concluding remarks summarizing the key findings and implications for future studies are given in Section \ref{sec4}.

\section{Model and method}
\label{sec2}

\begin{figure}
\begin{center}
\resizebox{\columnwidth}{!}{\includegraphics{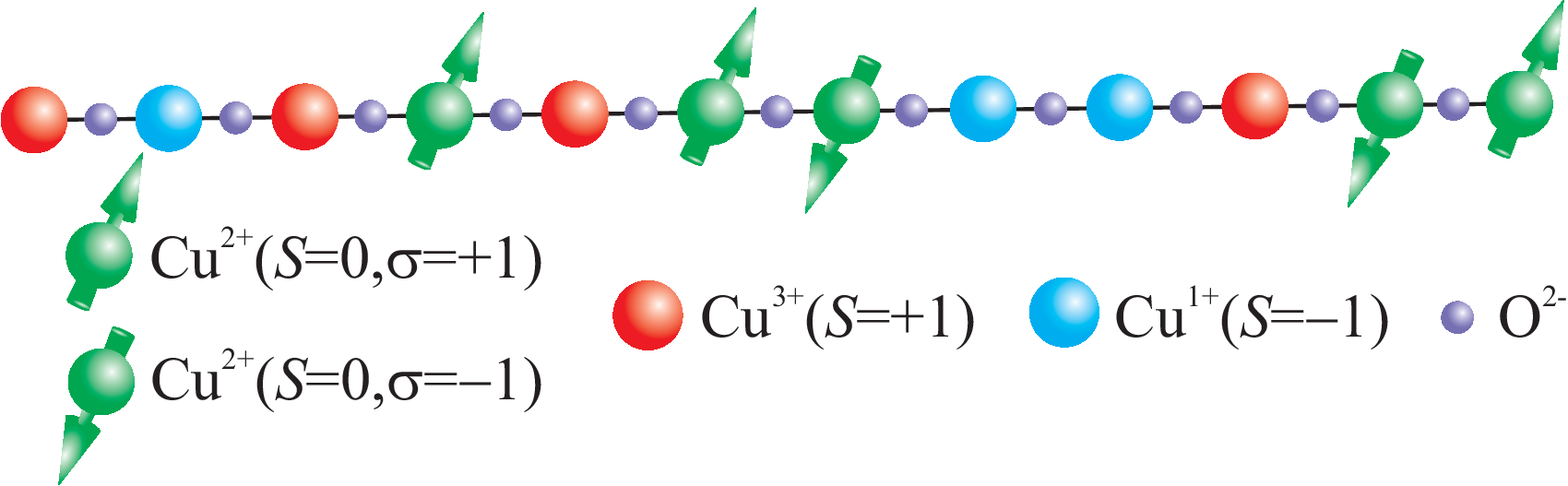}}
\end{center}
\vspace{-0.3cm}
\caption{An exemplary configuration of the considered spin-pseudospin model of the one-dimensional cuprate chain [CuO]$_\infty$. Large green spheres represent Cu$^{2+}$ centers (i.e., the electroneutral bare centers [CuO]$^0$) characterized by the pseudospin value $S=0$, whereby up- and down-pointing arrows specify a magnetic polarization characterized by the spin values $\sigma=1$ and $-1$, respectively. Large red and blue spheres represent Cu$^{3+}$ and Cu$^{1+}$ centers (i.e., positively charged hole centers [CuO]$^{+}$ and negatively charged electron centers [CuO]$^{-}$) characterized by the pseudospin values $S=1$  and $-1$, respectively. Small violet spheres correspond to 
O$^{2-}$ anions.}
\label{figmod}       
\end{figure}

Let us consider the minimal spin-pseudospin model of one-dimensional cuprate chain [CuO]$_\infty$, which takes into consideration three valence states of Cu$^{1+}$, Cu$^{2+}$ and Cu$^{3+}$ centers described by the pseudospin triplet with the projections $S = -1, 0, +1$, respectively. Within the polymeric cuprate chain [CuO]$_\infty$, the three nominal valence states of Cu$^{1+}$, Cu$^{2+}$ and Cu$^{3+}$ centers surrounded by O$^{2-}$ anions give rise to three distinct valence states of the [CuO] center: a negatively charged electron center [CuO]$^{-}$, an electroneutral bare center [CuO]$^0$, and a positively charged hole center [CuO]$^{+}$. While two charged pseudospin states [CuO]$^{-}$ and [CuO]$^{+}$ involve nonmagnetic Cu$^{1+}$ and Cu$^{3+}$ centers, respectively, the electroneutral pseudospin state [CuO]$^0$ encompasses a magnetic spin-1/2 Cu$^{2+}$ center characterized by the Ising spin $\sigma=\pm 1$. The non-magnetic charged states of [CuO]$^{-}$ and [CuO]$^{+}$ centers (nominally Cu$^{1+}$ and Cu$^{3+}$) are subject to electrostatic forces, which are absent in the electroneutral bare centers [CuO]$^{0}$ subjected solely to short-range magnetic forces. An exemplary configuration of the spin-pseudospin model of the one-dimensional cuprate chain [CuO]$_\infty$ is illustrated in Fig. \ref{figmod}. Bearing all this in mind, the minimal spin-pseudospin model of the one-dimensional cuprate chain [CuO]$_\infty$ can be described by the following effective Hamiltonian:
\begin{eqnarray}
{\cal H}\!=\! J \! \sum_{i=1}^N (1\!-\!S_i^2)\sigma_i(1\!-\!S_{i+1}^2)\sigma_{i+1}
\!+\!\Delta \! \sum_{i=1}^N \! S_{i}^2 \! + \! V \! \sum_{i=1}^N \! S_iS_{i+1}, \nonumber \\
\label{ham}
\end{eqnarray}
which is expressed in terms of the pseudospin triplet $S_i=-1,0,1$ corresponding to the three valence states of Cu$^{1+}$, Cu$^{2+}$, Cu$^{3+}$ centers and the Ising spin $\sigma_i=\pm 1$ ascribed to the magnetic Cu$^{2+}$ center only (i.e. $S_i=0$). The first term of the Hamiltonian (\ref{ham}) accounts for the antiferromagnetic exchange interaction $J>0$ between nearest-neighbor Cu$^{2+}$ centers, whereby $(1\!-\!S_i^2)$ acts as the respective projection operator for the magnetic pseudospin state $S_i=0$. The second term of the Hamiltonian (\ref{ham}) represents the on-site energy cost $\Delta>0$ associated with the creation of either a negatively charged electron center [CuO]$^{-}$ or a positively charged hole center [CuO]$^{+}$ corresponding to the nominal valence states Cu$^{1+}$ and Cu$^{3+}$, respectively. The third term of the Hamiltonian (\ref{ham}) takes into account the inter-site electrostatic energy between two nearest-neighbor charged centers, which lowers (raises) the overall electrostatic energy provided that two neighboring centers are oppositely (equally) charged. It should be noted that we impose electroneutrality constraints on the cuprate chain [CuO]$_\infty$, which means that the nominal valence states Cu$^{1+}$ and Cu$^{3+}$ leading to negatively [CuO]$^{-}$ and positively [CuO]$^{+}$ charged centers may only arise in pairs as a result of one-electron charge transfer: Cu$^{2+}$ + Cu$^{2+}$ $\to$ Cu$^{1+}$ + Cu$^{3+}$. 

The total Hamiltonian (\ref{ham}) of the spin-pseudospin model can be written as a sum over bond Hamiltonians:
\begin{eqnarray}
{\cal H}=\sum_{i=1}^N{\cal H}_i,
\end{eqnarray}
where the $i$-th bond Hamiltonian can be cast into the following symmetrized form:
\begin{eqnarray}
{\cal H}_i\!=\!J(1\!-\!S_i^2)\sigma_i(1\!-\!S_{i+1}^2)\sigma_{i+1}
\!+\!\frac{\Delta}{2}(S_{i}^2\!+\!S_{i+1}^2)\!+\! VS_iS_{i+1}. \nonumber \\
\label{hami}
\end{eqnarray}
The partition function of the spin-pseudospin model can be factorized into the following product:
\begin{eqnarray}
{\cal Z} \!&=&\! \sum_{\{S\}}\sum_{\{\sigma\}}\prod_{i=1}^N \frac{1}{2^{S_i^2}} {\rm e}^{-\beta {\cal H}_i}
=\sum_{\{S\}}\sum_{\{\sigma\}}\prod_{i=1}^N \frac{1}{\sqrt{2^{S_i^2+S_{i+1}^2}}} {\rm e}^{-\beta {\cal H}_i} \nonumber \\
\!&=&\! \sum_{\{S\}}\sum_{\{\sigma\}}\prod_{i=1}^N{\rm T}(S_i,\sigma_i;S_{i+1},\sigma_{i+1}), 
\label{pf}
\end{eqnarray}
where $\beta = 1/(k_{\rm B} T)$, $k_{\rm B}$ is Boltzmann's constant, $T$ is absolute temperature and the summations $\sum_{\{S\}}$ and $\sum_{\{\sigma\}}$ are carried out over all pseudospin and spin variables, respectively. The factor $\frac{1}{2^{S_i^2}}$ and its symmetrized counterpart $\frac{1}{\sqrt{2^{S_i^2 \!+\!S_{i+1}^2}}}$ are introduced as correction factors for two charged states $S_i = \pm 1$ to avoid their double counting in the partition function when performing the summation over the Ising spins $\sum_{\{\sigma\}}$. The Boltzmann factor ${\rm T}(S_i,\sigma_i;S_{i+1},\sigma_{i+1})=\frac{1}{\sqrt{2^{S_i^2 \!+\!S_{i+1}^2}}}{\rm e}^{-\beta {\cal H}_i}$ depending on the pseudospin and spin variables from two adjacent sites can be alternatively viewed as the transfer matrix:
\begin{eqnarray}
{\rm T} \!\!\!\!&&\!\!\!\!  (S_i,\sigma_i;S_{i+1},\sigma_{i+1}) \!=\! \frac{1}{\sqrt{2^{S_i^2 \!+\!S_{i+1}^2}}} 
\exp\left[-\frac{\beta\Delta}{2}(S_{i}^2+S_{i+1}^2)\right]\nonumber \\
\!&\times&\! \exp\left[-\beta V S_i S_{i+1} \!-\!\beta J(1\!-\!S_i^2)\sigma_i(1\!-\!S_{i+1}^2)\sigma_{i+1}\right].
\end{eqnarray}
The pseudospin and spin variables $S_i$ and $\sigma_i$ from $i$-th site define rows of the transfer matrix, while the pseudospin and spin variables $S_{i+1}$ and $\sigma_{i+1}$ from $(i+1)$-st site define columns of the transfer matrix, which has the following explicit form:
\begin{strip}
\rule[-1ex]{\columnwidth}{1pt}\rule[-1ex]{1pt}{1.5ex}
\begin{eqnarray}
{\rm T}&=&\left(
\begin{array}{cccccc}
{\rm T}(1,1;1,1) & {\rm T}(1,1;1,-1) & {\rm T}(1,1;0,1)& {\rm T}(1,1;0,-1) & {\rm T}(1,1;-1,1) & {\rm T}(1,1;-1,-1) \\ 	
{\rm T}(1,-1;1,1) & {\rm T}(1,-1;1,-1)& {\rm T}(1,-1;0,1) & {\rm T}(1,-1;0,-1) & {\rm T}(1,-1;-1,1) &{\rm T}(1,-1;-1,-1) \\
{\rm T}(0,1;1,1) & {\rm T}(0,1;1,-1) & {\rm T}(0,1;0,1) & {\rm T}(0,1;0,-1) &{\rm T}(0,1;-1,1)  & {\rm T}(0,1;-1,-1)\\
{\rm T}(0,-1;1,1) & {\rm T}(0,-1;1,-1) & {\rm T}(0,-1;0,1) & {\rm T}(0,-1;0,-1) & {\rm T}(0,-1;-1,1) &{\rm T}(0,-1;-1,-1)\\
{\rm T}(-1,1;1,1) & {\rm T}(-1,1;1,-1) & {\rm T}(-1,1;0,1) & {\rm T}(-1,1;0,-1) & {\rm T}(-1,1;-1,1) &{\rm T}(-1,1;-1,-1)\\
{\rm T}(-1,-1;1,1) & {\rm T}(-1,-1;1,-1) & {\rm T}(-1,-1;0,1) & {\rm T}(-1,-1;0,-1) & {\rm T}(-1,-1;-1,1) &{\rm T}(-1,-1;-1,-1)\\
\end{array} \right) \nonumber \\
&=&\left(
\begin{array}{cccccc}
\frac{1}{2}{\rm e}^{-\beta(\Delta+V)} & \frac{1}{2}{\rm e}^{-\beta(\Delta+V)} & \frac{1}{\sqrt{2}}{\rm e}^{-\frac{\beta \Delta}{2}}& \frac{1}{\sqrt{2}}{\rm e}^{-\frac{\beta \Delta}{2}} & \frac{1}{2}{\rm e}^{-\beta(\Delta-V)} & \frac{1}{2}{\rm e}^{-\beta(\Delta-V)} \\ 	
\frac{1}{2}{\rm e}^{-\beta(\Delta+V)} & \frac{1}{2}{\rm e}^{-\beta(\Delta+V)} & \frac{1}{\sqrt{2}}{\rm e}^{-\frac{\beta \Delta}{2}} & \frac{1}{\sqrt{2}}{\rm e}^{-\frac{\beta \Delta}{2}} &\frac{1}{2}{\rm e}^{-\beta(\Delta-V)}&\frac{1}{2}{\rm e}^{-\beta(\Delta-V)}  \\
\frac{1}{\sqrt{2}}{\rm e}^{-\frac{\beta \Delta}{2}} & \frac{1}{\sqrt{2}}{\rm e}^{-\frac{\beta \Delta}{2}} & {\rm e}^{-\beta J} & {\rm e}^{\beta J} &\frac{1}{\sqrt{2}}{\rm e}^{-\frac{\beta \Delta}{2}}  & \frac{1}{\sqrt{2}}{\rm e}^{-\frac{\beta \Delta}{2}}\\
\frac{1}{\sqrt{2}}{\rm e}^{-\frac{\beta \Delta}{2}}  &\frac{1}{\sqrt{2}}{\rm e}^{-\frac{\beta \Delta}{2}}  & {\rm e}^{\beta J} & {\rm e}^{-\beta J} & \frac{1}{\sqrt{2}}{\rm e}^{-\frac{\beta \Delta}{2}}  &\frac{1}{\sqrt{2}}{\rm e}^{-\frac{\beta \Delta}{2}} \\
 \frac{1}{2}{\rm e}^{-\beta(\Delta-V)}&  \frac{1}{2}{\rm e}^{-\beta(\Delta-V)} & \frac{1}{\sqrt{2}}{\rm e}^{-\frac{\beta \Delta}{2}} & \frac{1}{\sqrt{2}}{\rm e}^{-\frac{\beta \Delta}{2}} & \frac{1}{2}{\rm e}^{-\beta(\Delta+V)} &\frac{1}{2}{\rm e}^{-\beta(\Delta+V)}\\
 \frac{1}{2}{\rm e}^{-\beta(\Delta-V)} &  \frac{1}{2}{\rm e}^{-\beta(\Delta-V)} & \frac{1}{\sqrt{2}}{\rm e}^{-\frac{\beta \Delta}{2}} & \frac{1}{\sqrt{2}}{\rm e}^{-\frac{\beta \Delta}{2}} &\frac{1}{2}{\rm e}^{-\beta(\Delta+V)} &\frac{1}{2}{\rm e}^{-\beta(\Delta+V)}\\
\end{array}
\right).
\label{transferMatrix}
\end{eqnarray}
\hfill\rule[1ex]{1pt}{1.5ex}\rule[2.3ex]{\columnwidth}{1pt}
\end{strip}%
After straightforward analytical diagonalization of the transfer matrix (\ref{transferMatrix}), one obtains all six eigenvalues:
\begin{eqnarray}
\lambda_{1,2} &=& 0, \nonumber \\
\lambda_{3}   &=& -2\sinh(\beta J), \nonumber \\
\lambda_{4}   &=& -2\exp(-\beta\Delta)\sinh(\beta V), \nonumber \\
\lambda_{5}   &=& \cosh(\beta J)\!+\!\exp(-\beta\Delta)\cosh(\beta V)\!-\!\!\sqrt{D}, \nonumber \\
\lambda_{6}   &=& \cosh(\beta J)\!+\!\exp(-\beta\Delta)\cosh(\beta V)\!+\!\!\sqrt{D}, 
\label{tme}
\end{eqnarray}
where $D = \left[\cosh(\beta J)-{\rm e}^{-\beta\Delta}\cosh(\beta V)\right]^2 + 4{\rm e}^{-\beta\Delta}$. In the spirit of the transfer-matrix method, the subsequent summation over the pseudospin and spin variables in Eq. (\ref{pf}) is equivalent to a matrix multiplication. Utilizing a trace invariance, the partition function of the spin-pseudospin model can be consequently calculated from the transfer-matrix eigenvalues (\ref{tme}): 
\begin{eqnarray}
{\cal Z} = \sum_{\sigma_1 = \pm 1} \sum_{S_1 = \pm 1,0} \!\!\!\! {\rm T} (S_1,\sigma_1;S_1,\sigma_1) = \mbox{Tr}\, {\rm T}^N = \sum_{j=1}^6 \lambda_j^N\!.
\end{eqnarray}
It can be readily proved that the largest transfer-matrix eigenvalue is $\lambda_{max} = \lambda_6$, and hence, the free energy of the spin-pseudospin model in the thermodynamic limit $N \to \infty$ can be expressed solely in terms of the largest transfer-matrix eigenvalue:
\begin{eqnarray}
F = - k_{\rm B}T \lim_{N  \to \infty} \ln {\cal Z} = - N k_{\rm B} T \ln \lambda_6.
\label{fed}
\end{eqnarray}
The exact result (\ref{fed}) derived for the free energy  allows straightforward derivation of local observables such as the mean value of the pseudospin $\langle S_i \rangle$, the mean value of the square of the pseudospin $\langle S_i^2 \rangle$, the pair correlation function between nearest-neighbor pseudospins $\langle S_i S_{i+1}\rangle$, and the pair correlation function between nearest-neighbor spins $\langle \sigma_i \sigma_{i+1}\rangle$. In addition, standard thermodynamic relations can be utilized to  derive exact results for other quantities such as the magnetic entropy and specific heat from the free energy (\ref{fed}). 

\section{Results and discussion}
\label{sec3}

\begin{figure}
\begin{center}
\resizebox{\columnwidth}{!}{\includegraphics{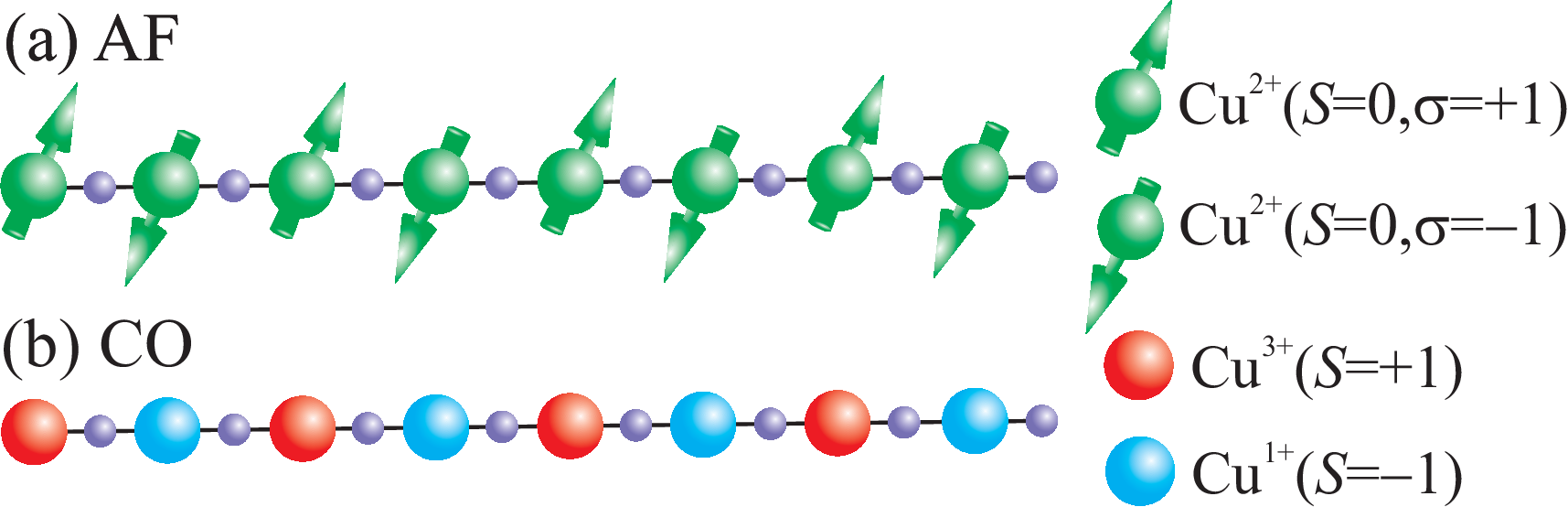}}
\end{center}
\vspace{-0.3cm}
\caption{A schematic illustration of two possible ground states of the spin-pseudospin model of the one-dimensional cuprate chain [CuO]$_\infty$: (a) the antiferromagnetic (AF) phase; (b) the charged ordered (CO) phase. The notation for the three nominal valence states of Cu$^{1+}$, Cu$^{2+}$, and Cu$^{3+}$ centers is the same as in Fig. \ref{figmod}.}
\label{figgs}       
\end{figure}

\begin{figure}
\begin{center}
\resizebox{0.9\columnwidth}{!}{\includegraphics{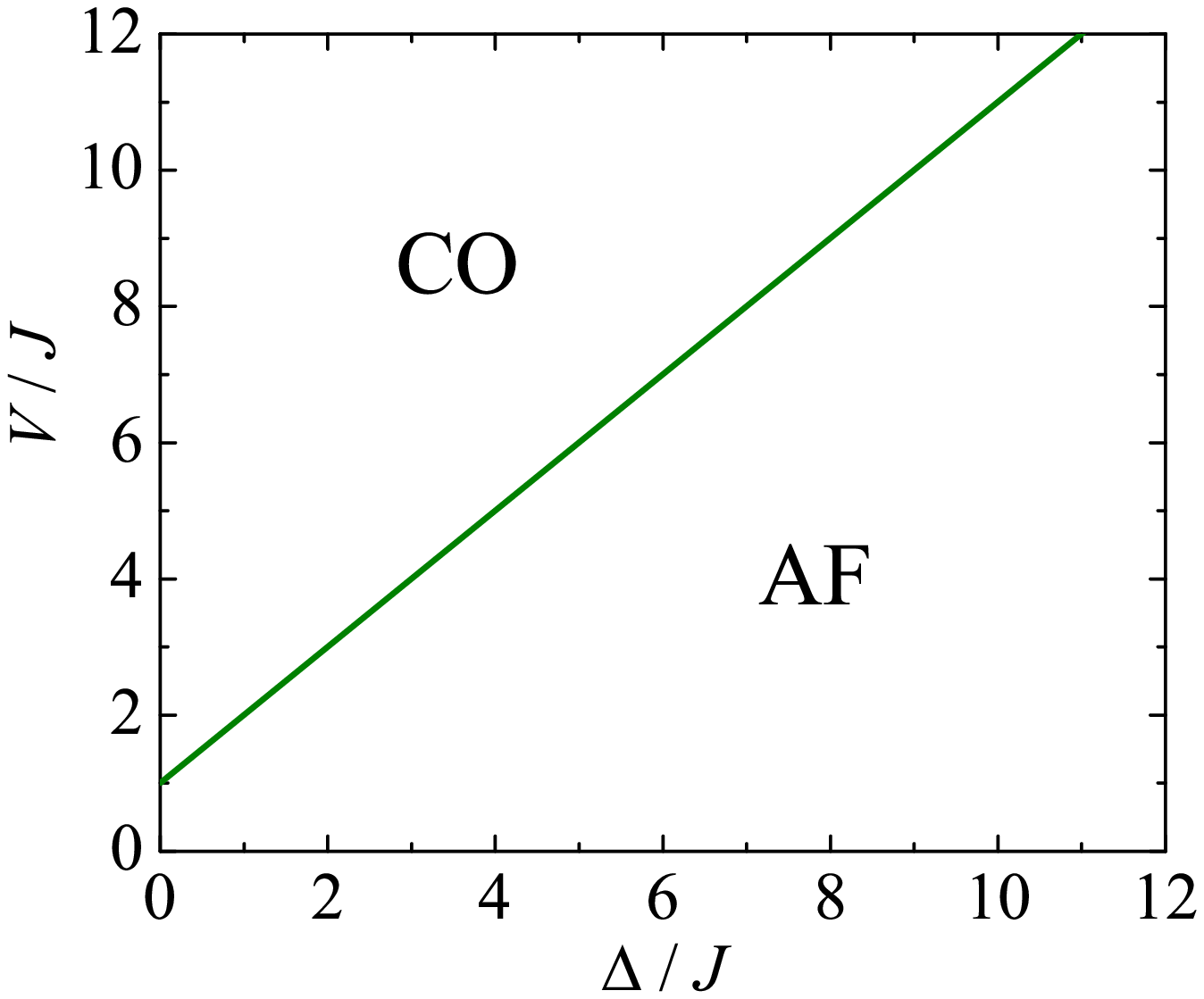}}
\end{center}
\vspace{-0.4cm}
\caption{The ground-state phase diagram of the spin-pseudospin model of the one-dimensional cuprate chain [CuO]$_\infty$ in the $\Delta/J-V/J$ plane. The antiferromagnetic (AF) and the charge ordered (CO) phases are separated by the line of first-order phase transitions $V/J = 1 + \Delta/J$.}
\label{fig1}       
\end{figure}

Let us proceed to a discussion of the most intriguing findings  obtained for the minimal pseudospin-spin model of the one-dimensional cuprate chain [CuO]$_\infty$. It turns out that the spin-pseudospin model exhibits only two ground states exemplified by configurations illustrated in Fig. \ref{figgs}. The first ground state comprises solely the electroneutral [CuO]$^{0}$ centers indicating a regular alternation of Cu$^{2+}$ and O$^{2-}$ ions within the entire one-dimensional cuprate chain [CuO]$_\infty$. Moreover, the spins of nearest-neighbor Cu$^{2+}$ centers display antiferromagnetic alignment characterizing this ground state as the antiferromagnetic (AF) phase. The AF phase evidently manifests spontaneously broken symmetry and consists of two-fold degenerate eigenvectors with the energy $E_{\rm AF}=-NJ$: 
\begin{eqnarray}
|{\rm AF}\rangle = \left \{ \begin{array}{l}
                         \displaystyle \prod_{i=1}^{N/2} |S=0, \sigma=+1 \rangle_{2i-1} |S=0, \sigma=-1 \rangle_{2i} \\
												 \displaystyle \prod_{i=1}^{N/2} |S=0, \sigma=-1 \rangle_{2i-1} |S=0, \sigma=+1 \rangle_{2i} 
                       \end{array}
							\right..			
\label{eq:qaf}
\end{eqnarray}
In the second ground state, one contrarily finds a regular alternation of negatively charged electron centers [CuO]$^{-}$ with positively charged hole centers [CuO]$^{+}$, which means that nonmagnetic Cu$^{1+}$ and Cu$^{3+}$ centers (separated by O$^{2-}$ anions) regularly alternate within the one-dimensional cuprate chain [CuO]$_\infty$. Consequently, the second ground state can be regarded as the charged ordered (CO) phase, which also exhibits spontaneously broken symmetry and consists of two-fold degenerate eigenvectors with the energy $E_{\rm CO}=-NV+N\Delta$:
\begin{eqnarray}
|{\rm CO}\rangle = \left \{ \begin{array}{l}
                         \displaystyle \prod_{i=1}^{N/2} |S=+1 \rangle_{2i-1} |S=-1 \rangle_{2i} \\
												 \displaystyle \prod_{i=1}^{N/2} |S=-1 \rangle_{2i-1} |S=+1 \rangle_{2i} 
                       \end{array}
							\right..			
\label{eq:co}
\end{eqnarray} 
It is noteworthy that the AF and CO phases coexist whenever the attractive inter-site interaction $V$ between the negatively charged electron centers [CuO]$^{-}$ (nominally Cu$^{1+}$ centers) and the positively charged hole centers [CuO]$^{+}$ (nominally Cu$^{3+}$ centers) is balanced by the antiferromagnetic superexchange interaction $J$ between the spin-1/2 Cu$^{2+}$ ions inherent to the electroneutral [CuO]$^{0}$ centers and the on-site energy cost associated with the creation of the charged centers. The corresponding line of first-order phase transitions $V = \Delta + J$ between the AF and CO phases is depicted in the ground-state phase diagram in Fig. \ref{fig1}. The most intriguing behavior of the pseudospin-spin model could be thus anticipated if the interaction parameters drive the one-dimensional cuprate chain [CuO]$_\infty$ close to a first-order phase transition between the AF and CO phases. 

\begin{figure}
\begin{center}
\resizebox{0.9\columnwidth}{!}{\includegraphics{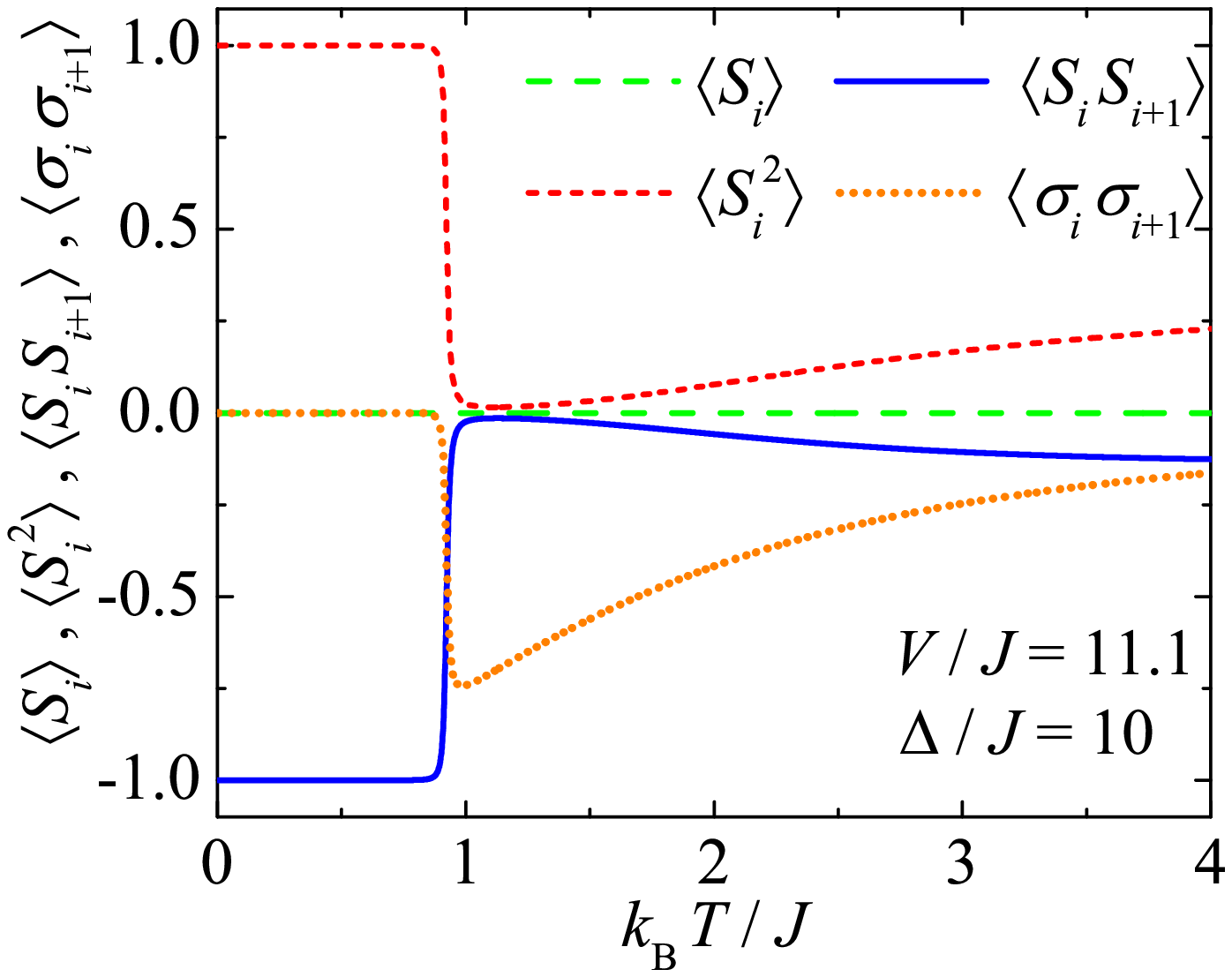}}
\end{center}
\vspace{-0.4cm}
\caption{Temperature variations of the mean value of pseudospin $\langle S_i\rangle$, the mean square value of pseudospin $\langle S_i^2\rangle$, the pair correlation function between nearest-neighbor pseudospins $\langle S_iS_{i+1}\rangle$, and the pair correlation function between nearest-neighbor spins $\langle \sigma_i\sigma_{i+1}\rangle$ of the spin-pseudospin model of the one-dimensional cuprate chain [CuO]$_\infty$ with fixed values of the interaction parameters $V/J=11.1$ and $\Delta/J=10$.}
\label{fig2}       
\end{figure}

The temperature variations of two local observables ($\langle S_i\rangle$, $\langle S_i^2\rangle$) and two pair correlation functions ($\langle S_i S_{i+1}\rangle$, $\langle \sigma_i \sigma_{i+1}\rangle$) are depicted in Fig. \ref{fig2} for the spin-pseudospin model of the one-dimensional cuprate chain [CuO]$_\infty$ with specific values of the interaction parameters $V/J=11.1$ and $\Delta/J=10$. Notably, the mean value of the pseudospin remains zero $\langle S_i\rangle=0$ irrespective of temperature, which convincingly evidences an equal proliferation of both charged centers [CuO]$^{-}$ and [CuO]$^{+}$ (nominally Cu$^{1+}$ and Cu$^{3+}$ centers), and hence, satisfying the electroneutrality condition. Besides, the mean square value of pseudospin tends towards its maximum value $\langle S_i^2\rangle=1$ in the zero-temperature limit in accordance with the CO ground state. This mean value remains nearly constant up to the temperature $k_{\rm B}T/J\approx 1$, at which it abruptly decreases almost to zero indicating the absence of charged centers. The abrupt breakdown of the mean square value of pseudospin $\langle S_i^2\rangle$ from unity to zero signifies the pseudo-transition from the CO phase towards the AF phase. A rather steep increase of the pair correlation function between nearest-neighbor pseudospins from the perfectly anticorrelated value $\langle S_i S_{i+1}\rangle=-1$ almost towards zero provides further evidence of the pseudo-transition from the CO phase towards the AF phase. The absence of a pair correlation function between nearest-neighbor spins $\langle \sigma_i \sigma_{i+1} \rangle = 0$ at absolute zero temperature also confirms the CO ground state. However, a steep decline towards the value $\langle \sigma_i\sigma_{i+1}\rangle\approx -0.75$ observed around the temperature $k_{\rm B}T/J\approx 1$ indicates the pseudo-transition towards the AF phase, where a perfectly anticorrelated value $\langle \sigma_i\sigma_{i+1}\rangle=-1$ would be expected at absolute zero temperature. 

\begin{figure}
\begin{center}
\resizebox{0.9\columnwidth}{!}{\includegraphics{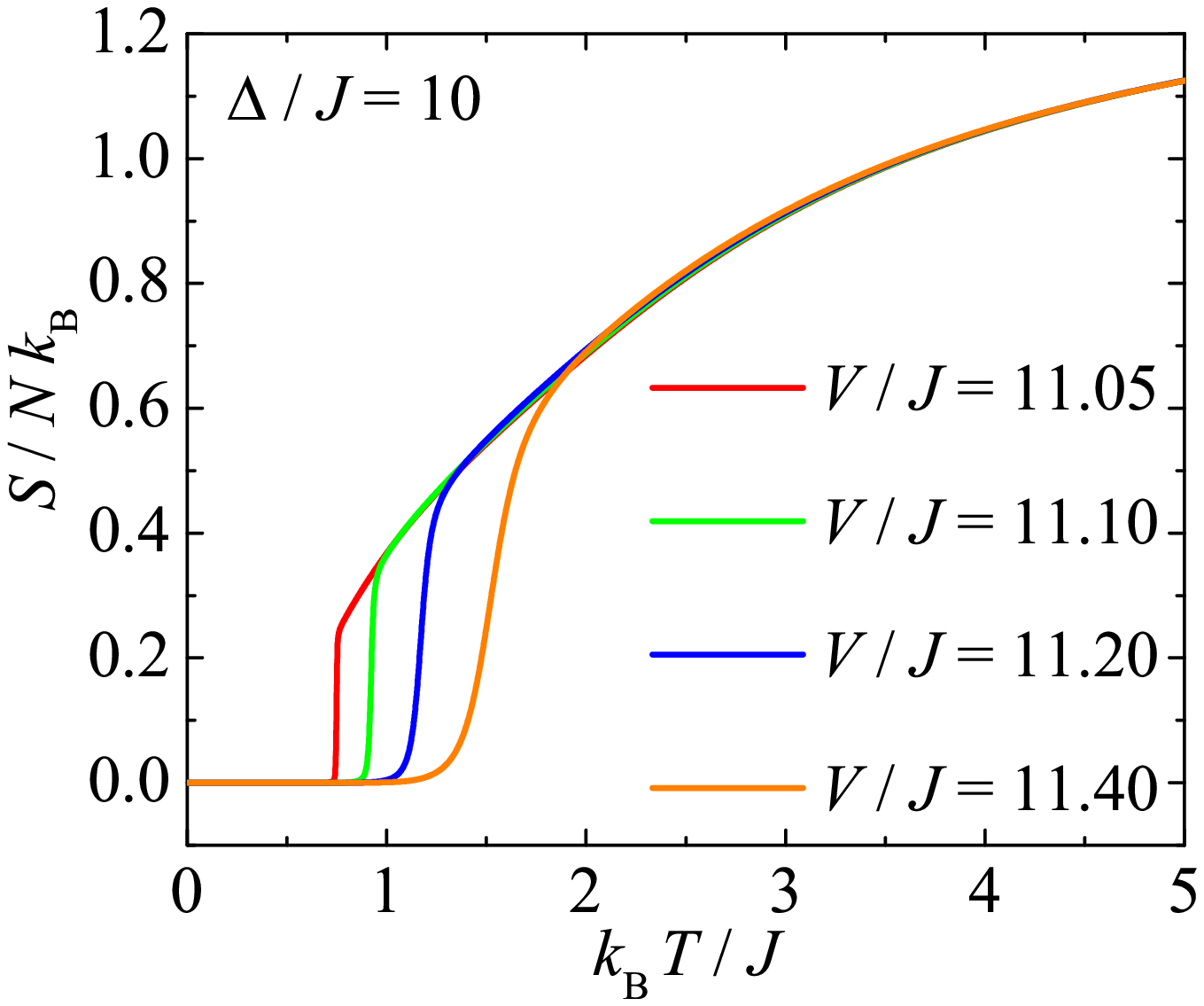}}
\end{center}
\vspace{-0.4cm}
\caption{Temperature dependencies of the entropy of the spin-pseudospin model of the one-dimensional cuprate chain [CuO]$_\infty$ for the fixed value of the on-site interaction $\Delta/J=10$ and a few different values of the inter-site interaction $V/J$.}
\label{figent}       
\end{figure}

Next, let us bring insight into how the pseudo-transition between the CO and AF phases manifests itself in the temperature dependencies of the magnetic entropy and specific heat. To this end, the temperature dependencies of the entropy of the spin-pseudospin model of the one-dimensional cuprate chain [CuO]$_\infty$ are depicted in Fig. \ref{figent} for a fixed value of the on-site interaction $\Delta/J=10$ and a few selected values of the inter-site interaction $V/J$. Generally, the entropy maintains zero value at sufficiently low temperatures, then it exhibits a sudden rise within a relatively narrow range of moderate temperatures before it finally shows a more gradual increase at higher temperatures. It is evident from Fig. \ref{figent} that the sudden rise in entropy becomes sharper as the inter-site interaction approaches the ground-state phase boundary between the CO and AF phases. Although the steep increase in entropy entropy resembles a discontinuous thermal phase transition, the observed increase in entropy still remains continuous. 

\begin{figure}
\begin{center}
\resizebox{0.9\columnwidth}{!}{\includegraphics{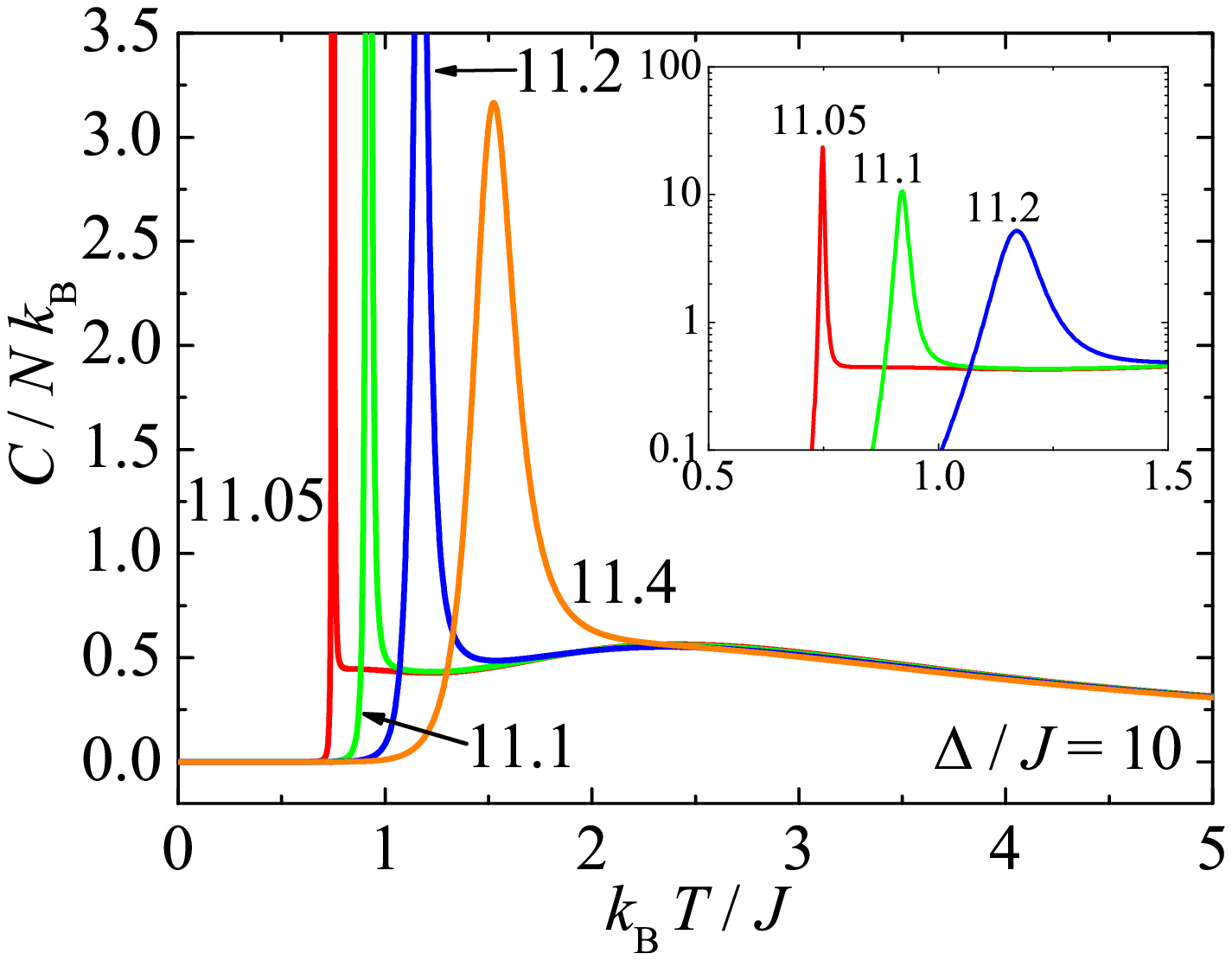}}
\end{center}
\vspace{-0.4cm}
\caption{Temperature dependencies of the specific heat of the spin-pseudospin model of one-dimensional cuprate chain [CuO]$_\infty$ for the fixed value the on-site interaction $\Delta/J=10$ and four selected values of the inter-site interaction $V/J = 11.05$, $11.1$, $11.2$, and $11.4$ specified by the respective labels. The inset shows in a semi-logarithmic scale detail from a temperature range, where a sizable but still finite maximum of the specific heat is observed.}
\label{figsh}       
\end{figure}

To complete the overall picture, temperature variations of the specific heat of the spin-pseudospin model of the one-dimensional cuprate chain [CuO]$_\infty$ are displayed in Fig. \ref{figsh} for the same fixed value of the on-site interaction $\Delta/J=10$ and a few different values of the inter-site interaction $V/J$ selected slightly above the ground-state phase boundary between the CO and AF phases ($V/J=11$ for $\Delta/J=10$). It is obvious from Fig. \ref{figsh} that the specific heat exhibits an extremely high peak at relatively low temperatures, which is thus quite reminiscent of a continuous thermal phase transition. However, it should be pointed out that the relevant peak of specific heat, although it may show increase over several orders of magnitude, always remains finite (see the insert of Fig. \ref{figsh}). If the inter-site interaction is chosen further away from the ground-state phase boundary between the CO and AF phases, the specific-heat peak shifts towards higher temperatures, reduces its height and becomes broader. 

\begin{figure}
\begin{center}
\resizebox{0.9\columnwidth}{!}{\includegraphics{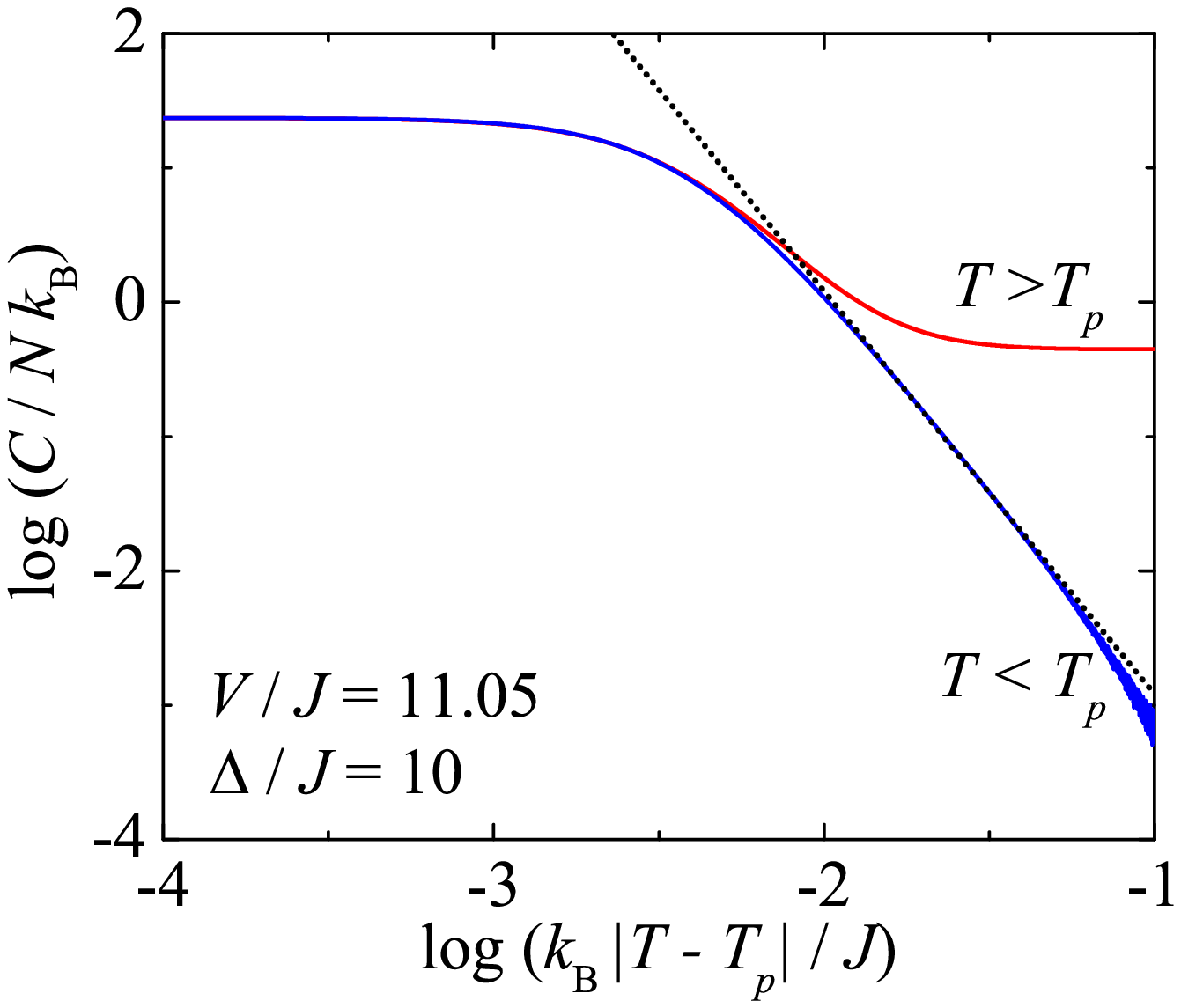}}
\end{center}
\vspace{-0.4cm}
\caption{Temperature dependence of the specific heat of the spin-pseudospin model of the one-dimensional cuprate chain [CuO]$_\infty$ close to the pseudotransition in a log-log scale for fixed values of the parameters $\Delta/J=10$ and $V/J = 11.05$. A blue (red) line displays the temperature dependence of the specific heat slightly below (above) of the pseudotransition temperature. A linear fit  of the specific heat $\log (C/Nk_{\rm B}) = -5.92 - 3 \log(k_{\rm B}|T - T_p|/J)$ shown by a black dotted line verifies a power-law dependence below the pseudotransition temperature.}
\label{figce}       
\end{figure}

Furthermore, it might be also interesting to explore the asymptotic behavior of the specific heat near the pseudo-critical temperature. To this end, we plotted in Fig. \ref{figce} in a log-log scale temperature dependence of the specific heat for the spin-pseudospin model of the one-dimensional cuprate chain [CuO]$_\infty$ with fixed values of the on-site interaction $\Delta/J=10$ and the inter-site interaction $V/J = 11.05$. Fig. \ref{figce} illustrates that the specific heat closely follows a power-law dependence with the respective quasi-critical exponent $\alpha' = 3$ slightly below the pseudo-critical temperature even though it eventually bends and converges to a finite value as the pseudo-critical temperature is approached. On the other hand, the relevant power-law dependence of the specific heat cannot be clearly discerned in a temperature range slightly above the pseudo-critical temperature where the specific heat rapidly drops to an almost constant nonzero value (see the insert in Fig. \ref{figsh}). Nonetheless, it can be inferred that the spin-pseudospin model of the one-dimensional cuprate chain [CuO]$_\infty$ falls within the universality class of one-dimensional models
exhibiting a pseudo-transition at finite temperatures \cite{roj19}.

\begin{figure}
\begin{center}
\resizebox{0.9\columnwidth}{!}{\includegraphics{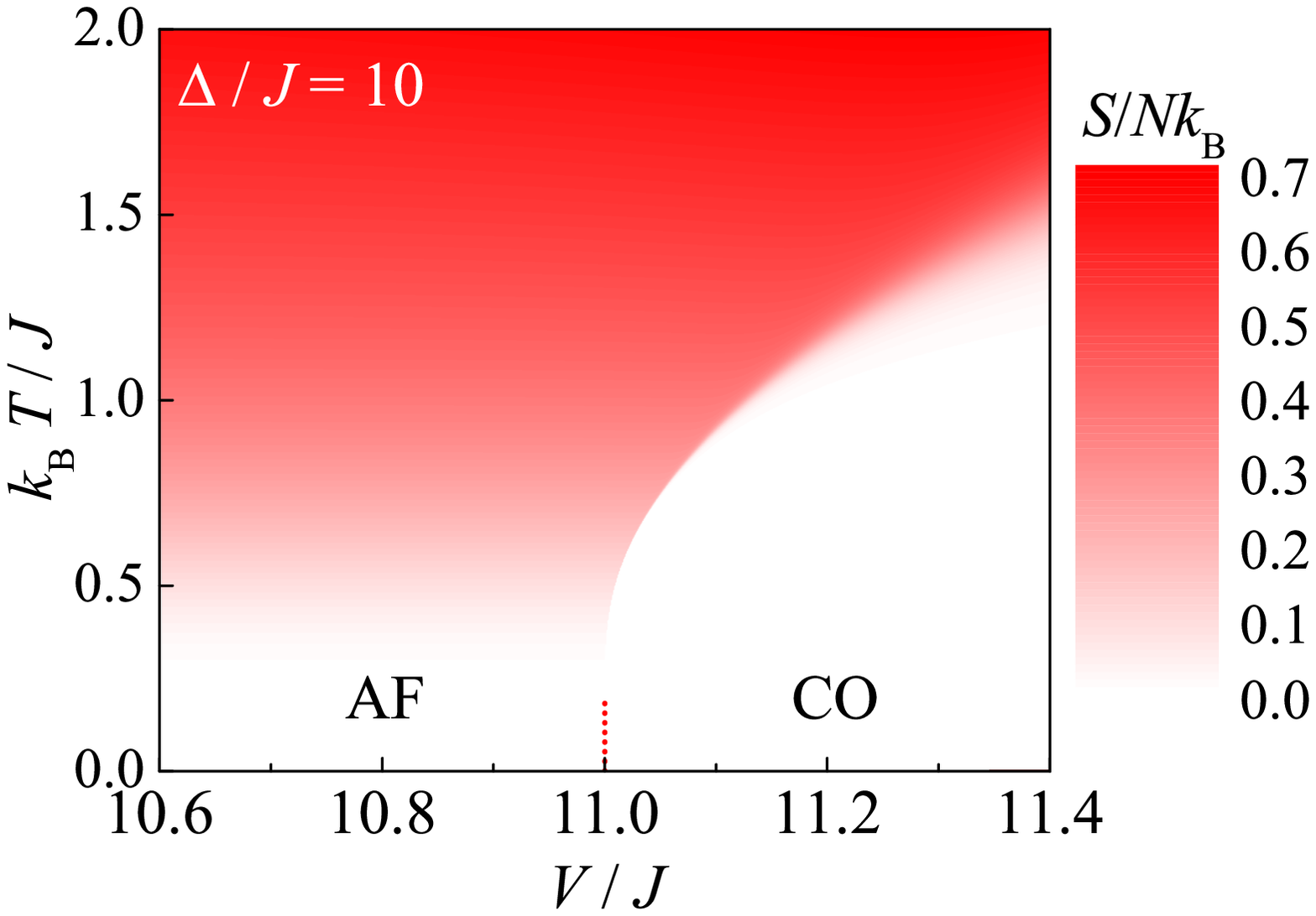}}
\end{center}
\vspace{-0.4cm}
\caption{A density plot of the entropy of the spin-pseudospin model of the one-dimensional cuprate chain [CuO]$_\infty$ in the $V/J-k_{\rm B} T/J$ plane for a fixed value of the parameter $\Delta/J=10$. The specific value $V/J = 11$ corresponds to the ground-state phase boundary between the AF and CO phases.}
\label{figde}       
\end{figure}

One can exploit the abrupt change in entropy and the extremely sharp peak of the specific heat, which accompany the pseudo-transition between the CO and AF phases, to construct the pseudo-phase diagrams of the spin-pseudospin model of the one-dimensional cuprate chain [CuO]$_\infty$. The density plot of the entropy shown in Fig. \ref{figde} in the inter-site interaction versus temperature ($V/J-k_{\rm B} T/J$) plane directly implies that the AF phase gains much higher entropy upon increasing  temperature compared to the CO phase. This finding stems from the fact that all excited charged configurations emerging above the CO ground state still have to satisfy the electroneutrality constraint with an equal number of positively and negatively charged centers [CuO]$^{-}$ and [CuO]$^{+}$ (nominally Cu$^{1+}$ and Cu$^{3+}$ centers), whereas all spin configurations emerging above the AF ground state naturally satisfy the electroneutrality constraint as they are composed of electroneutral centers [CuO]$^{0}$ (nominally Cu$^{2+}$). Owing to this fact, the line of pseudo-phase transitions between the AF and CO phases bends towards the CO phase due to entropic favoring of the AF phase.

\begin{figure}
\begin{center}
\resizebox{0.9\columnwidth}{!}{\includegraphics{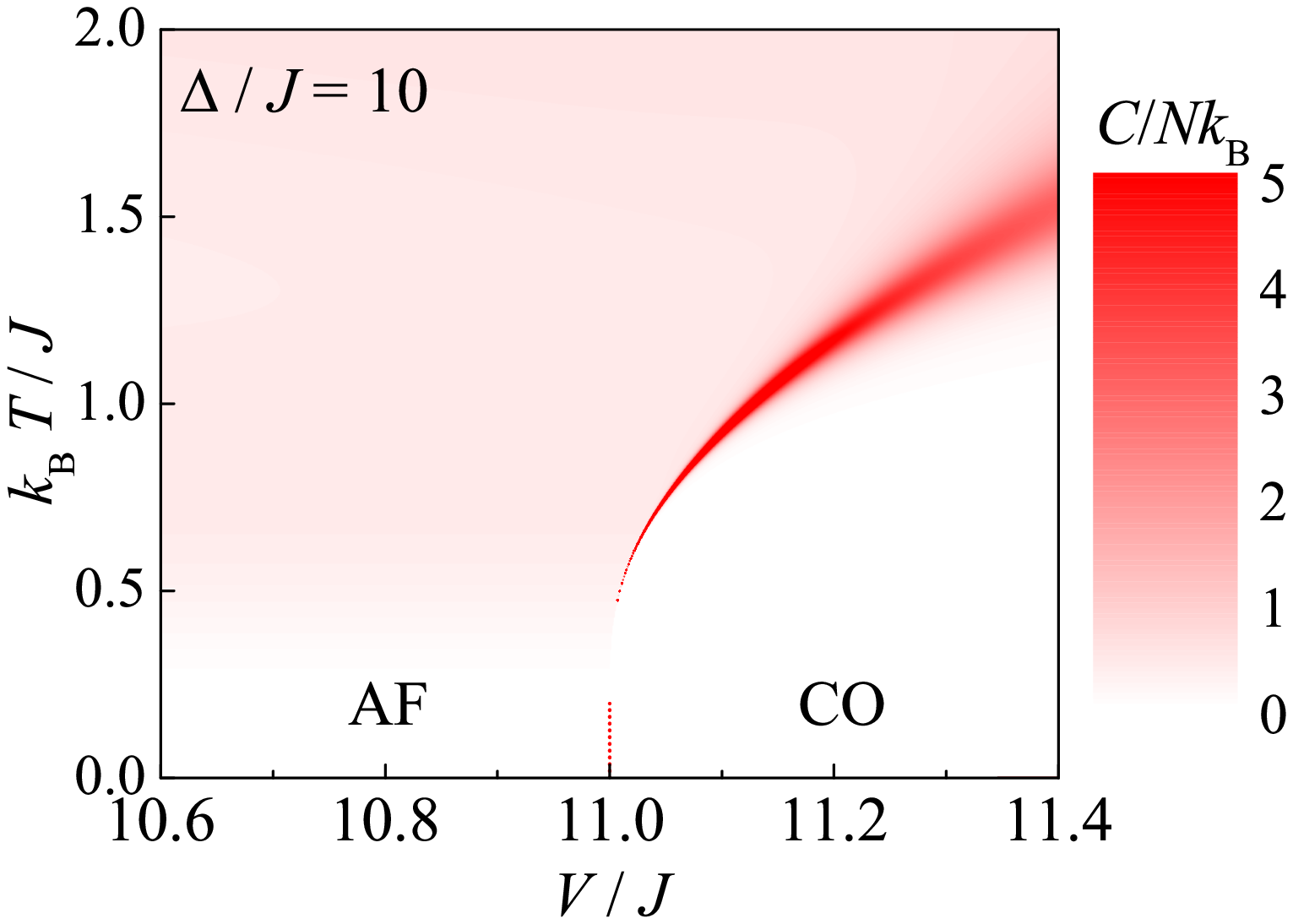}}
\end{center}
\vspace{-0.4cm}
\caption{A density plot of the specific heat of the spin-pseudospin model of the one-dimensional cuprate chain [CuO]$_\infty$ in the $V/J-k_{\rm B} T/J$ plane for a fixed value of the parameter $\Delta/J=10$. The specific value $V/J = 11$ corresponds to the ground-state phase boundary between the AF and CO phases.}
\label{figdsh}       
\end{figure}

The density plot of the specific heat presented in Fig. \ref{figdsh} in the inter-site interaction versus temperature ($V/J-k_{\rm B} T/J$) provides another clear pseudo-phase diagram of the spin-pseudospin model of the one-dimensional cuprate chain [CuO]$_\infty$. It is evident from Fig. \ref{figdsh} that the pseudo-phase boundary between the CO and AF phases is almost perfectly vertical up to moderate temperatures $k_{\rm B} T/J \approx 0.4$ before it begins to bend towards the CO phase due to the entropically-driven favoring of the AF phase. It should also be stressed that the pseudo-phase transition between the CO and AF phases gradually melts upon increasing temperature. It actually turns out that the vigorous peak of specific heat observed in a very narrow temperature range persists up to moderate temperatures $k_{\rm B} T/J \approx 1.0$, while at higher temperatures, the relevant peak of specific heat is gradually suppressed and spreads over a wider temperature range.
 
\section{Conclusion}
\label{sec4}

In the present article, we have addressed the minimal spin-pseudospin model of the one-dimensional cuprate chain [CuO]$_\infty$, which takes into consideration the negatively charged electron center [CuO]$^{-}$, the electroneutral center [CuO]$^{0}$, and positively charged hole center [CuO]$^{+}$ corresponding to three nominal valence states of Cu$^{1+}$, Cu$^{2+}$, and Cu$^{3+}$ ions, respectively. The transfer-matrix solution of the model under investigation afforded exact results for the free energy, entropy, specific heat, as well as, local order parameters and pair correlation functions. It has been demonstrated that the investigated spin-pseudospin model has two available ground states, which can be characterized as the AF and CO phases. Although the AF and CO phases do not exhibit true spontaneous long-range order, the low-temperature behavior of local observables and pair correlation functions clearly verifies characteristic short-range arrangement of the AF and CO phases.

The most intriguing discovery highlighted in this study concerns with the manifestation of a pseudo-transition between the CO and AF phases. This phenomenon emerges when the spin-pseudospin model of the one-dimensional cuprate chain [CuO]$_\infty$ approaches the CO ground state near the ground-state phase boundary with the AF phase. Under these conditions, the entropy exhibits a sudden thermally-induced increase reminiscent of a discontinuous phase transition, while the specific heat displays a prominent sharp peak reminiscent of a continuous phase transition. Despite the absence of a true jump in entropy jump and divergence in specific heat, all aforedescribed features serve as clear signatures of the pseudo-transition. For future investigations, it would be valuable to relax the electroneutrality condition to explore  the effects of electron or hole doping on the pseudo-transition phenomenon. In addition, it would be intriguing to investigate whether similar pseudo-transition phenomena can be found in more complex low-dimensional cuprates such ladders or plane square lattices. 

\begin{acknowledgement}
This work was financially supported by the grant of the Slovak Research and Development Agency provided under the contract Nos. APVV-16-0186 and APVV-20-0150 and by the grant of The Ministry of Education, Science, Research, and Sport of the Slovak Republic provided under the contract No. VEGA 1/0695/23. This project has received funding from the European Union's Horizon 2020 research and innovation programme under the Marie Sk\l odowska-Curie grant agreement No. 945380.
\end{acknowledgement}

\section*{Author contributions}
Both authors contributed equally to the paper.

\section*{Conflict of interest}
 The authors have no conflicts of interest to disclose.

\section*{Data availability statements}
This manuscript has no associated data or the data will not be deposited [Authors’ comment: This is a theoretical study with no experimental data. All plots directly follow from exact analytical formulas included in the manuscript.].


\begin{thebibliography}{50}

\bibitem{mat93} 
D. C. Mattis, The Many-Body Problem: An Encyclopedia of Exactly Solved Models in One Dimension. World Scientific, Singapore (1993).

\bibitem{hov50} 
L. van Hove, Sur L'int\'egrale de Configuration Pour Les Syst\'emes  De Particules A Une Dimension (On the Complete Configuration of One-dimensional Particle Systems).
Physica \textbf{16}, 137 (1950). doi: \url{https://doi.org/10.1016/0031-8914(50)90072-3}

\bibitem{cue04}
J. A. Cuesta, A. S\'anchez, General non-existence theorem for phase transitions in one-dimensional systems with short range interactions, and physical examples of such transitions. 
J. Stat. Phys. \textbf{115}, 869 (2004). doi: \url{https://doi.org/10.1023/B:JOSS.0000022373.63640.4e}

\bibitem{gal15} 
L. G\'alisov\'a, J. Stre\v{c}ka, Vigorous thermal excitations in a double-tetrahedral chain of localized Ising spins and mobile electrons mimic a temperature-driven first-order phase transition. 
Phys. Rev. E, \textbf{91}, 0222134 (2015). doi: \url{https://doi.org/10.1103/PhysRevE.91.022134}

\bibitem{roj16}
O. Rojas, J. Stre\v{c}ka, S. M. de Souza, Thermal entanglement and sharp specific-heat peak in an exactly solved spin-1/2 Ising-Heisenberg ladder with alternating Ising and Heisenberg inter-leg couplings. 
Solid State Commun., \textbf{246}, 68-75 (2016). doi: \url{https://doi.org/10.1016/j.ssc.2016.08.002}

\bibitem{str16} 
J. Stre\v{c}ka, R.C. Al\'ecio, M.L . Lyra, O. Rojas, 
Spin frustration of a spin-1/2 Ising-Heisenberg three-leg tube as an indispensable ground for thermal entanglement. 
J. Magn. Magn. Mater., \textbf{409}, 124-133 (2016). doi: \url{https://doi.org/10.1016/j.jmmm.2016.02.095}

\bibitem{car18} 
I. M. Carvalho, J. Torrico, S. M. de Souza, M. Rojas, O. Rojas, 
Quantum entanglement in the neighborhood of pseudo-transition for a spin-1/2 Ising-XYZ diamond chain. 
J. Magn. Magn. Mater., \textbf{465}, 323 (2018). doi: \url{https://doi.org/10.1016/j.jmmm.2018.06.018}

\bibitem{car19} 
I. M. Carvalho, J. Torrico, S. M. de Souza, O. Rojas, O. Derzhko, 
Correlation functions for a spin-1/2 Ising-XYZ diamond chain: Further evidence for quasi-phases and pseudo-transitions. 
Ann. Phys., \textbf{402}, 45-65 (2019). doi: \url{https://doi.org/10.1016/j.aop.2019.01.001}

\bibitem{ono20} 
O. Rojas, Residual Entropy and Low Temperature Pseudo-Transition for One-Dimensional Models. 
Acta Phys. Pol. A, \textbf{137}, 933 (2020). doi: \url{https://doi.org/10.12693/APhysPolA.137.933}

\bibitem{roj20}
O. Rojas, J. Stre\v{c}ka, O. Derzhko, S.M. de Souza, Peculiarities in pseudo-transitions of a mixed spin-(1/2,1) Ising-Heisenberg double-tetrahedral chain in an external magnetic field. 
J. Phys.: Condens. Matter, \textbf{32}, 035804 (2020). doi: \url{https://doi.org/10.1088/1361-648X/ab4acc}

\bibitem{str20} 
J. Stre\v{c}ka, Anomalous thermodynamic response in the vicinity of a pseudo-transition of a spin-1/2 Ising diamond chain. 
Acta Phys. Pol. A, \textbf{137}, 610-612 (2020). doi: \url{https://doi.org/10.12693/APhysPolA.137.610}

\bibitem{nov20}
J. Stre\v{c}ka, Pseudo-Critical Behavior of Spin-1/2 Ising Diamond and Tetrahedral Chains, in: An Introduction to the Ising Model, Ed. S. Luoma, Nova Science Publishers, New York, 2020, pp. 63-86, Chapter 4.

\bibitem{pim22} 
R. A. Pimenta, O. Rojas, S.M. de Souza, Anomalous thermodynamics in a mixed spin-1/2 and spin-1 hexagonal nanowire system.
J. Magn. Magn. Mater., \textbf{550}, 169070 (2022). doi: \url{https://doi.org/10.1016/j.jmmm.2022.169070}

\bibitem{hut21} 
T. Hutak, T. Krokhmalskii, O. Rojas, S. M. de Souza, O. Derzhko, 
Low-temperature thermodynamics of the two-leg ladder Ising model with trimer rungs: A mystery explained. 
Phys. Lett. A, \textbf{387}, 127020 (2021). doi: \url{https://doi.org/10.1016/j.physleta.2020.127020}

\bibitem{yin24} 
W. Yin, Paradigm for approaching the forbidden spontaneous phase transition in the one-dimensional Ising model at a fixed finite temperature. 
Phys. Rev. Res., \textbf{6}, 013331 (2024). doi: \url{https://doi.org/10.1103/PhysRevResearch.6.013331}

\bibitem{pan21}
Y. Panov, O. Rojas, Unconventional low-temperature features in the one-dimensional frustrated q-state Potts model.
Phys. Rev. E, \textbf{103}, 062107 (2021). doi: \url{https://doi.org/10.1103/PhysRevE.103.062107}

\bibitem{pan23} 
Y. Panov, O. Rojas, 
Zero-temperature phase transitions and their anomalous influence on thermodynamic behavior in the q-state Potts model on a diamond chain.
Phys. Rev. E, \textbf{108}, 044144 (2023). doi: \url{https://doi.org/10.1103/PhysRevE.108.044144}

\bibitem{sou18}
S. M. de Souza, O. Rojas, Quasi-phases and pseudo-transitions in one-dimensional models with nearest neighbor interactions.
Solid State Commun., \textbf{269}, 131-134 (2018). doi: \url{https://doi.org/10.1016/j.ssc.2017.10.006}

\bibitem{oro20}
O. Rojas, 
A conjecture on the relationship between critical residual entropy and finite temperature pseudo-transitions of one-dimensional models. 
Braz. J. Phys., \textbf{50}, 675 (2020). doi: \url{https://doi.org/10.1007/s13538-020-00773-8}

\bibitem{roj19}
O. Rojas, J. Stre\v{c}ka, M. L. Lyra, S. M. de Souza, Universality and quasicritical exponents of one-dimensional models displaying a quasitransition at finite temperatures. 
Phys. Rev. E, \textbf{99}, 042117 (2019). doi: \url{https://doi.org/10.1103/PhysRevE.99.042117}

\bibitem{kro21} 
T. Krokhmalskii, T. Hutak, O. Rojas, S. M. de Souza, O. Derzhko, Towards low-temperature peculiarities of thermodynamic quantities for decorated spin chains.
Physica A, \textbf{573}, 125986 (2021). doi: \url{https://doi.org/10.1016/j.physa.2021.125986}

\bibitem{roj21}
O. Rojas, Finite size effects around pseudo-transition in one-dimensional models with nearest neighbor interaction. 
Chinese J. Phys. \textbf{70}, 157 (2021). doi: \url{https://doi.org/10.1016/j.cjph.2021.01.002}

\bibitem{mos11} 
A. S. Moskvin, True charge-transfer gap in parent insulating cuprates. 
Phys. Rev. B \textbf{84}, 075116 (2011). doi: \url{https://doi.org/10.1103/PhysRevB.84.075116}

\bibitem{mos13}
A. S. Moskvin, Perspectives of disproportionation driven superconductivity in strongly correlated 3d compounds. 
J. Phys.: Condens. Matter \textbf{25}, 085601 (2013). doi: \url{https://doi.org/10.1088/0953-8984/25/8/085601}

\bibitem{mos19}
A. S. Moskvin, Yu. D. Panov, Topological Structures in Unconventional Scenario for 2D Cuprates.
J. Supercond. Nov. Magn. \textbf{32}, 61 (2019). doi: \url{https://doi.org/10.1007/s10948-018-4896-0}

\bibitem{pan16} 
Yu. D. Panov, A. S. Moskvin, A. A. Chikov, I. L. Avvakumov, Competition of spin and charge orders in a model cuprate.
J. Supercond. Nov. Magn. \textbf{29}, 1077 (2016). doi: \url{https://doi.org/10.1007/s10948-016-3378-5}

\bibitem{pan17}
Yu. D. Panov, A. S. Moskvin, A. A. Chikov, K. S. Budrin, The ground-state phase diagram of 2D spinpseudospin system.
J. Low Temp. Phys. \textbf{187}, 646 (2017). doi: \url{https://doi.org/10.1007/s10909-017-1743-9}

\bibitem{pan19} 
Yu. D. Panov, V. A. Ulitko, K. S. Burdin, A. A. Chikov, A. S. Moskvin, Phase diagrams of a 2D Ising spin-pseudospin model.
J. Magn. Magn. Mater. \textbf{477}, 162 (2019). doi: \url{https://doi.org/10.1016/j.jmmm.2019.01.049}

\bibitem{pan20} 
D. N. Yasinskaya, V. A. Ulitko, A. A. Chikov, Yu. D. Panov, Critical Behavior of a 2D Spin-Pseudospin Model in a Strong Exchange Limit.
Acta Phys. Pol. A \textbf{137}, 979 (2020). doi: \url{https://doi.org/10.12693/APhysPolA.137.979}

\bibitem{pan22}
A. S. Moskvin, Yu. D. Panov, Model of charge triplets for high-$T_c$ cuprates.
J. Magn. Magn. Mater. \textbf{550}, 169004 (2022). doi: \url{https://doi.org/10.1016/j.jmmm.2021.169004}

\bibitem{ami95} 
T. Ami, M. K. Crawford, R. L. Harlow, Z. R. Wang, D. C. Johnston, Q. Huang, R. W. Erwin, 
Magnetic susceptibility and low-temperature structure of the linear chain cuprate Sr$_2$CuO$_3$.
Phys. Rev. B \textbf{51}, 5994 (1995). doi: \url{https://doi.org/10.1103/PhysRevB.51.5994}

\bibitem{mai97}
K. Maiti, D. D. Sarma, T. Mizokawa, A. Fujimori, Electronic structure of one-dimensional cuprate Sr$_2$CuO$_3$.
Europhys. Lett. \textbf{37}, 359 (1997). doi: \url{ https://doi.org/10.1209/epl/i1997-00157-x}

\bibitem{sch12} 
J. Schlappa, K. Wohlfeld, K. Zhou \textit{et al}., Spin-orbital separation in the quasi-one-dimensional Mott insulator Sr$_2$CuO$_3$.
Nature \textbf{485}, 82 (2012). doi: \url{https://doi.org/10.1038/nature10974}

\bibitem{sch18}
J. Schlappa, U. Kumar, K. J. Zhou \textit{et al}., 
Probing multi-spinon excitations outside of the two-spinon continuum in the antiferromagnetic spin chain cuprate Sr$_2$CuO$_3$.
Nat. Commun. \textbf{9}, 5394 (2018). doi: \url{https://doi.org/10.1038/s41467-018-07838-y}

\bibitem{yam95} 
K. Yamada, J. Wada, S. Hosoya, Y. Endoh, S. Noguchi, S. Kawamata, K. Okuda, 
Antiferromagnetic long range order of the S=1/2 linear chain cuprate Ca$_2$CuO$_3$.
Physica C \textbf{253}, 135-138 (1995). doi: \url{https://doi.org/10.1016/0921-4534(95)00503-X}

\bibitem{nag97} 
N. Nagasako, T. Oguchi, H. Fujisawa, O. Akaki, T. Yokoya, T. Takahashi, M. Tanaka, M. Hasegawa, H. Takei, 
Electronic Band Structure of SrCuO$_2$.
J. Phys. Soc. Jpn. \textbf{66}, 1756-1761 (1997). doi: \url{https://doi.org/10.1143/JPSJ.66.1756}

\bibitem{bou18} 
D. Bounoua, R. Saint-Martin, S. Petit, F. Bourdarot, L. Pinsard-Gaudart, 
Finite size effect on the magnetic excitations spectra, phonons and heat conduction of the quasi-one-dimensional spin chains system SrCuO$_2$.
Physica B \textbf{536}, 323 (2018). doi: \url{https://doi.org/10.1016/j.physb.2017.10.104}

\end{thebibliography}
\end{document}